\documentclass[10pt,a4paper,twocolumn]{article}
\usepackage[footnotesize]{caption}
\usepackage[english]{babel}
\usepackage[margin={0.575in,0.8in}]{geometry}
\usepackage{times,amsmath,epsfig,multirow,booktabs,amsthm,amsmath,amsfonts,amssymb}
\usepackage[caption=false,font=footnotesize]{subfig}
\newcommand{\eps}{\varepsilon}
\newtheorem{thm}{Theorem}
\newtheorem{defn}{Definition}
\newtheorem{lem}{Lemma}
\newtheorem{cor}{Corollary}
\newtheorem{ex}{Example}

\title{Dynamic Traitor Tracing Schemes, Revisited}
\author{Thijs Laarhoven\footnote{T. Laarhoven is with the Department of Mathematics and Computer Science, Eindhoven University of Technology, The Netherlands. \protect\\
E-mail: mail@thijs.com.}}
\date{\today}

\begin{document}
\maketitle

\begin{abstract}
We revisit recent results from the area of collusion-resistant traitor tracing, and show how they can be combined and improved to obtain more efficient dynamic traitor tracing schemes. In particular, we show how the dynamic Tardos scheme of Laarhoven et al.\ can be combined with the optimized score functions of Oosterwijk et al.\ to trace coalitions much faster. If the attack strategy is known, in many cases the order of the code length goes down from quadratic to linear in the number of colluders, while if the attack is not known, we show how the interleaving defense may be used to catch all colluders about twice as fast as in the dynamic Tardos scheme. Some of these results also apply to the static traitor tracing setting where the attack strategy is known in advance, and to group testing.
\end{abstract}


\section{Introduction}
\label{sec:intro}

Suppose a distributor of streaming digital content wishes to sell his content to customers. To do this, he may encrypt the stream and grant paying customers access to the live stream to prevent others from freely accessing the data. However, a malicious customer may try to broadcast the stream directly to non-paying users by creating a pirate broadcast, possibly making money out of it as well. To prevent this, the distributor may embed unique watermarks into the content for each user, such that when a pirate streams his copy of the content online, the distributor may detect it, extract the watermark from the content, and see who is the pirate. 

Things become more complicated when several pirates \textit{collude} and mix their copies of the content into a new stream, with a sequence of watermarks that is not equal to any of their individual sequences of watermarks. Under the assumption that in each segment, these colluders can only output one of their watermarked segments (known as the marking assumption), it is still possible to trace collusions of any size $c$, by carefully choosing which watermarked segments each user should get, and designing an algorithm that traces (the watermark of) the pirate stream to the set of colluders. In the setting of live streams, the assignment of watermarks for the next content segments may be adjusted based on the previous pirate output; we call this the \textit{dynamic} traitor tracing setting. The scenario where the assignment of watermarks to users has to be done completely in advance, and where users may only be disconnected from the stream at the end, is often referred to as \textit{static} traitor tracing. The latter scenario may for instance apply to video on demand. 

The goal of collusion-resistant traitor tracing is to use as few watermarked content segments as possible, while still being able to trace the users guilty of creating the pirate broadcast.

\subsection{Model}

The above dynamic traitor tracing problem is often translated to the following mathematical model. First, we assume there is a set of $n$ users, $c$ of which are colluding to create the pirate output. We assume that in each content segment, the distributor is able to embed up to $q$ different watermarks, with $q \ll n$ often referred to as the alphabet size. To each segment and each user, the distributor assigns one of $q$ watermarks. This assignment is denoted by a code matrix $X = (X_{j,i})$, where $X_{j,i} = k$ means that in segment $i$, user $j$ receives the $k$th (out of $q$) watermark. 

After distributing the $i$th watermarked segments to all users, the set of colluders chooses one of their segments and outputs it, along with its watermark. We denote the sequence of watermarks in the pirate output by a vector $y = (y_i)$. Then, the distributor detects $y_i$, and uses this information to possibly disconnect some users that he thinks are guilty. After this, he chooses the next content segments $X_{j,i+1}$, possibly based on the previous pirate output $y_i$. This continues until no pirates are left to output one of their segments, or until we have reached the limit of the number of watermarked segments. We say a scheme is successful if, within an a priori fixed amount of segments $\ell$, it is able to guarantee (with high probability) that all colluders are caught, and no innocent users are accidentally disconnected along the way. We write $\eps_2$ for the probability of not catching all colluders, and $\eps_1$ for the probability of ever disconnecting at least one innocent user.

\subsection{Related work}

Fiat and Tassa~\cite{fiat01} showed that if the alphabet size $q$ satisfies $q \geq 2c + 1$, then with $\ell = O(c \log n)$ segments, one can find and disconnect all colluders deterministically, i.e., with probability of error $\eps_1 = \eps_2 = 0$. This scheme is very efficient, and the only drawback is that it requires a large alphabet size, which may not be possible in practice. Tassa~\cite{tassa05} later showed that combining the binary ($q = 2$) static scheme of Boneh and Shaw~\cite{boneh98} with the dynamic scheme of Fiat and Tassa~\cite{fiat01} leads to a binary dynamic traitor tracing scheme that uses $\ell = O(c^4 \log n)$ content segments. Due to the large required length of the code, this scheme is not practical.

Recently, Laarhoven et al.~\cite{laarhoven13tit} showed that the celebrated binary static scheme of Tardos~\cite{tardos03} can be converted into a dynamic binary scheme, with a code length of the order $\ell = O(c^2 \log n)$. Together with the divide-and-conquer construction of Laarhoven et al.~\cite{laarhoven12wifs}, allowing for a linear trade-off between the alphabet size and the code length, this allows one to build $q$-ary schemes with a code length of the order $\ell = O(\frac{c^2}{q} \log n)$, for arbitrary values of $q$. For large $q \approx O(c)$, this approaches the result of Fiat and Tassa, both in the alphabet size and in the number of segments needed.

Even more recently, Oosterwijk et al.~\cite{oosterwijk13} studied the score function used in Tardos' static scheme, and showed how to choose better score functions when the pirate attack is known. In one particular case, they came across a score function that turned out to work well against \textit{any} attack. More precisely, for asymptotically large $c$, they showed that this score function achieves capacity, i.e., attains the known exact lower bound on the code length of $\ell \sim 2 c^2 \ln n$ in the binary setting~\cite{huang12}. So for large $c$, using this new score function in the static setting may lead to a decrease in the length of the code of almost $60\%$ compared to using the symmetric score function of \v{S}kori\'{c} et al.~\cite{skoric08}, for which it is known that the asymptotic code length $\ell \sim \frac{1}{2} \pi^2 c^2 \ln n \approx 4.93 c^2 \ln n$ is optimal~\cite{laarhoven13ihmmsec}.

\subsection{Contributions and outline}

The outline of this paper is as follows. First, in Section~\ref{sec:tardos} we will revisit the dynamic Tardos scheme, and show how slightly modifying the scheme leads to a simpler proof and a slightly improved performance. Then, in Section~\ref{sec:scores} we take a closer look at the recent results of Oosterwijk et al.\ and show when and how these results can be improved and combined with the construction of the dynamic Tardos scheme to obtain efficient dynamic traitor tracing schemes. More precisely: we get a greatly improved performance when the attack strategy is completely known, as the order of the code length often goes down from quadratic in $c$ to linear in $c$; and we heuristically show how to get the above-mentioned decrease in the asymptotic code length of up to $60\%$ in dynamic traitor tracing, when no assumptions are made on the pirate attack. Finally, in Section~\ref{sec:conclusion}, we summarize the results and mention some directions for future work. 

In this paper we restrict our attention to binary, dynamic schemes, although some of the results may also be generalized to arbitrary alphabets, or applied to static traitor tracing.


\section{The Dynamic Tardos Scheme, Revisited}
\label{sec:tardos}

Let us first recall the dynamic Tardos traitor tracing scheme of Laarhoven et al.~\cite{laarhoven13tit}, outlined below. The main difference between the celebrated static Tardos scheme~\cite{tardos03} and the dynamic version of this scheme, is that while in the static scheme we disconnect a user only when his final score exceeds a certain threshold $Z$, in the dynamic scheme we disconnect users as soon as their scores exceed $Z$. With this small adjustment, it can be shown that whereas in the static Tardos scheme \textit{at least one} pirate is caught with high probability, in the dynamic scheme \textit{all} pirates are caught with high probability.

The construction below depends on scheme parameters $\ell$, $Z$, and $\delta$ (a cutoff parameter on the distribution function $F_{\delta}$). Choosing these parameters appropriately depends on whether one wants provable security or heuristic security. The provable bounds on $\eps_1$ and $\eps_2$ are not very tight, as the actual error probabilities are often significantly smaller. Heuristically, one may therefore be able to use shorter code lengths $\ell$, at the risk of not satisfying provable upper bounds. In this paper we will not go into detail about choosing $\ell$, $Z$, and $\delta$; see e.g. \cite{laarhoven13tit}, \cite{blayer08,laarhoven12dcc,
skoric13} for theoretical upper bounds, and \cite{furon09,simone12} for practical estimates of $\ell$.

\begin{enumerate}
	\item \textbf{Codeword generation} \\
	For each position $1 \leq i \leq \ell$:
	\begin{enumerate}
		\item Select $p_i \in [\delta, 1 - \delta]$ from $F_{\delta}(p)$ defined as:
		\begin{align}
			F_{\delta}(p) = \frac{2 \arcsin(\sqrt{p}) - 2 \arcsin(\sqrt{\delta})}{\pi - 4 \arcsin(\sqrt{\delta})}\, . \label{dist1}
		\end{align}
		\item Generate $X_{j,i} \in \{0,1\}$ using $P(X_{j,i} = 1) = p_i$.
	\end{enumerate}
	\item \textbf{Distribution/Detection/Accusation} \\
	For each position $1 \leq i \leq \ell$:
	\begin{enumerate}
		\item Send to each active user $j$ symbol $X_{j,i}$.
		\item Detect the output $y_i$; terminate if there is none.
		\item Calculate scores $S_{j,i} = h(X_{j,i},y_i,p_i)$ using:
	\end{enumerate}
		\begin{align}
		h(x,y,p) = \begin{cases}
			+\sqrt{p/(1 - p)}, & (x,y) = (0,0), \\
			-\sqrt{p/(1 - p)}, & (x,y) = (0,1), \\
			-\sqrt{(1 - p)/p}, & (x,y) = (1,0), \\
			+\sqrt{(1 - p)/p}, & (x,y) = (1,1).
		\end{cases} \label{scoresym}
		\end{align}
	\begin{enumerate}
		\item[d)] For active users $j$, set $S_j(i) = S_j(i-1) + S_{j,i}$, and disconnect this user if $S_j(i) > Z$.
	\end{enumerate}
\end{enumerate}

By making the following small modification, one can get a slight improvement in the theoretical bounds on $\ell$ of \cite{laarhoven13tit} and get a simplified proof. The main difference is that scores are only updated in segments where no one is disconnected.

\begin{enumerate}
	\itemindent0.7cm\item[2d)] If there are active users with $S_j(i-1) + S_{j,i} > Z$:
	\begin{itemize}
		\itemindent0.7cm\item Disconnect these users. 
		\itemindent0.7cm\item For other active users, set $S_j(i) = S_j(i-1)$.
	\end{itemize}
	\hspace{0.7cm}Else set $S_j(i) = S_j(i-1) + S_{j,i}$ for active users $j$. 
\end{enumerate}

Due to lack of space we omit the details, but with this modification, the proof will not have to resort to using the technical parameter $\tilde{Z}$ which is used in the proofs in \cite{laarhoven13tit}. It now suffices to prove that in positions where scores were calculated, the average pirate score will exceed $Z$. Both error probabilities $\eps_1$, $\eps_2$ still increase by (at most) a factor $2$, and the code length increases by an additive term $c$, but otherwise we can use the same parameters as in the static Tardos scheme. This modification thus leads to a cleaner reduction between code lengths in the static ($\ell^{(s)}$) and dynamic ($\ell^{(d)}$) Tardos schemes as follows, with notation similar to \cite{laarhoven12wifs}:
\begin{align}
\ell^{(d)}(c,n,\eps_1,\eps_2) = \ell^{(s)}(c,n,\tfrac{1}{2}\eps_1,\tfrac{1}{2}\eps_2) + c.
\end{align}
To give an indication of the provable code lengths with and without this modification: in the example of \cite{laarhoven13tit} with scheme parameters $c = 25$, $n = 10^6$, $\eps_{1,2} = 10^{-3}$, the number of segments now increases from $109\,585$ (static) to only $114\,204$ (dynamic), instead of the $116\,561$ of \cite{laarhoven13tit}.


\section{Optimal Score Functions, Revisited}
\label{sec:scores}

Recently, Oosterwijk et al.~\cite{oosterwijk13} showed how to obtain optimized score functions for any attack strategy, maximizing the normalized expected coalition score $\tilde{\mu}$ per segment\footnote{Here, normalized refers to dividing by the standard deviation of innocent user scores per segment, which we will denote by $\sigma$.}. For several attacks, they showed that this leads to much more efficient schemes, sometimes reducing the order of the code length from $c^2$ to as little as $c^{1.5}$ when using the arcsine distribution for generating biases $p_i$. As a side result, they noticed that the score function built against the interleaving attack (randomly choosing a pirate, and outputting his watermarked segment) performs well against any attack. 

In this section, we will discuss these improved score functions, as well as the optimal distribution functions against these attacks, in several different settings. We first discuss the scenario where the complete attack strategy is either known to the distributor in advance or can be accurately determined along the way~\cite{furon09b}, and how drastically the code lengths may be reduced in this setting. Then we will discuss the scenario where the attack strategy is not known and may not even be the same in each segment, in which case we show how we may be able to use the interleaving defense to our advantage. 


\subsection{Known attacks}

If the attack strategy is completely known, we can not only use the optimized score functions of Oosterwijk et al.~\cite{oosterwijk13} to increase the performance, but also fix the value of $p$ in advance to its optimal value. After all, if the attack strategy is known and identical for every column, we can compute the performance of the scheme as a function of $p$, and choose the value of $p$ that maximizes it. Note that using different values of $p$ for different columns is required in the ordinary traitor tracing setting, as an adversary could otherwise exploit this knowledge about $p$ to build strong attacks. But if we already know what the attack looks like in advance, then the model should be considered a randomized model rather than an adversarial model, and varying $p$ is no longer necessary.

Table~\ref{tab:1} lists some common attacks also discussed in \cite{oosterwijk13} and the optimal choices of $h$ and $p$ for each of these attacks, together with the asymptotic scaling of the lengths of the codes in each of these cases. Except for against the interleaving attack, in each of these examples the order of the number of segments needed to trace collusions goes down from $O(c^2 \log n)$ to $O(c \log n)$, thus leading to a drastic reduction in the code length. Below we briefly discuss each of these attacks, and give some intuition behind the results in the table. Due to space limitations, details on the optimal values of $p$, $h$, and the asymptotics of $\ell$ are omitted here but will appear in the full version of \cite{laarhoven13allerton}.

\begin{table*}[t]
\centering

    \caption{Optimal parameter choices for several attack strategies. The plots on the left show $P(y = 1 \mid k)$ against $k$, where $k \in \{0, \dots, c\}$ is the number of ones received by the coalition, when the corresponding attacks are used by the coalition.}
    \label{tab:1}
\renewcommand{\arraystretch}{1.3}
    \begin{small}
    \begin{tabular}{p{2.8cm}p{8.2cm}p{2.2cm}p{3cm}}
    \toprule
    {\bfseries Attack} & {\bfseries Optimal score function $h$} & {\bfseries Optimal bias $p$} & {\bfseries Asymptotics $\ell$} \\ 
    \midrule
    Interleaving attack & \multirow{4}{*}{$h(x,y) = \begin{cases} 
	+p/(1 - p) \qquad \qquad \qquad \quad \	\	& (x,y) = (0,0) \\
	-1 											& (x,y) = (0,1) \\
	-1 											& (x,y) = (1,0) \\ 
	+(1 - p)/p									& (x,y) = (1,1) 
	\end{cases}$} & \multirow{2}{*}{$p = \dfrac{1}{2}$}  & \multirow{2}{*}{$\ell \sim 2 c^2 \ln n$} \\
    \multirow{3}{*}{\includegraphics[width=2.5cm]{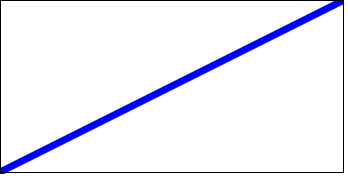}}	& &	& \\
            & &	\multirow{2}{*}{$\Big(p \in (0,1),$} & \multirow{2}{*}{$\ell \sim 2 c^2 \ln n\Big)$} \\
            & & & \\
    \midrule
	All-$1$ attack & \multirow{4}{*}{$h(x,y) = \begin{cases} 
	+p/(1 - p) 									& (x,y) = (0,0) \\
	-p(1 - p)^{c - 1}/(1 - (1 - p)^c) 			& (x,y) = (0,1) \\
	-\infty 									& (x,y) = (1,0) \\ 
	+(1 - p)^c/(1 - (1 - p)^c)					& (x,y) = (1,1) 
	\end{cases}$} & \multirow{2}{*}{$p = o\left(\dfrac{1}{c}\right)$} & \multirow{2}{*}{$\ell \sim 2 c \ln n$} \\
	\multirow{3}{*}{\includegraphics[width=2.5cm]{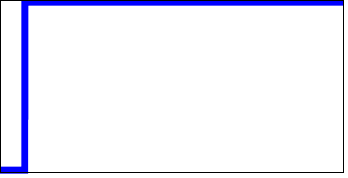}}			& & & \\
& & \multirow{2}{*}{$\Big(p = \dfrac{1}{c}\, ,$} & \multirow{2}{*}{$\ell \sim \dfrac{2e(e-1)}{2e-1} c \ln n\Big)$} \\
			& & & \\
    \midrule
    Coin-flip attack & $h(x,y) = \left(p^{c-1} + (1 - p)^{c-1}\right)$ & \multirow{2}{*}{$p = o\left(\dfrac{1}{c}\right)$} & \multirow{2}{*}{$\ell \sim 4 c \ln n$} \\
	\multirow{4}{*}{\includegraphics[width=2.5cm]{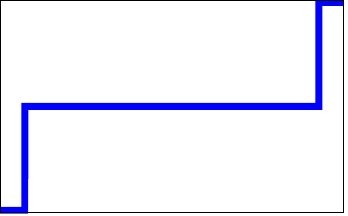}}	& \multirow{4}{*}{$\qquad \times \begin{cases}
	+p/(1 - p^c + (1 - p)^c) \qquad \quad \ \	& (x,y) = (0,0) \\ 
	-p/(1 + p^c - (1 - p)^c)					& (x,y) = (0,1) \\ 
	-(1 - p)/(1 - p^c + (1 - p)^c)				& (x,y) = (1,0) \\ 
	+(1 - p)/(1 + p^c - (1 - p)^c)				& (x,y) = (1,1) \\ 
	\end{cases}$} & & \\ 
& & \multirow{2}{*}{$\Big(p = \dfrac{1}{c}\, ,$} & \multirow{2}{*}{$\ell \sim 2(e^2-1) c \ln n\Big)$} \\
		& & & \\
		& & & \\
    \midrule
    \hyphenation{voting}
    Majority attack & \multirow{4}{*}{$h(x,y) = \begin{cases} 
	+1 \qquad \qquad \qquad \qquad \qquad \quad	& (x,y) = (0,0) \\
	-1 											& (x,y) = (0,1) \\
	-1 											& (x,y) = (1,0) \\ 
	+1											& (x,y) = (1,1) 
	\end{cases}$} &	\multirow{2}{*}{$p = \dfrac{1}{2}$} & \multirow{2}{*}{$\ell \sim \pi c \ln n$} \\
	\multirow{3}{*}{\includegraphics[width=2.5cm]{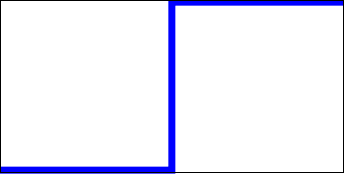}} & & & \\
		& & & \\
		& & & \\
    \midrule
    Minority attack & \multirow{4}{*}{$h(x,y) \approx \begin{cases} 
	+p/(1 - p) 									& (x,y) = (0,0) \\
	-p(1 - p)^{c - 1}/(1 - (1 - p)^c) 			& (x,y) = (0,1) \\
	-1 											& (x,y) = (1,0) \\ 
	+(1 - p)^c/(1 - (1 - p)^c)					& (x,y) = (1,1) 
	\end{cases}$} &	\multirow{2}{*}{$p = o\left(\dfrac{1}{c}\right)$} & \multirow{2}{*}{$\ell \sim 2 c \ln n$} \\
	\multirow{3}{*}{\includegraphics[width=2.5cm]{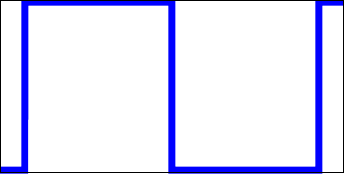}}		& & & \\ 
		& & \multirow{2}{*}{$\Big(p = \dfrac{1}{c}\, ,$} & \multirow{2}{*}{$\ell \sim 2(e-1) c \ln n\Big)$} \\
		& & & \\
    \midrule
    Unknown attacks & \multirow{4}{*}{$h(x,y,p) = \begin{cases} 
	+p/(1 - p) \qquad \qquad \qquad \ 	 		& (x,y) = (0,0) \\
	-1 											& (x,y) = (0,1) \\
	-1 											& (x,y) = (1,0) \\ 
	+(1 - p)/p									& (x,y) = (1,1) 
	\end{cases}$} &	\multirow{2}{*}{$p \sim F_{\delta}$} & \multirow{2}{*}{$\ell \lesssim 2 c^2 \ln n$} \\
	\multirow{3}{*}{\includegraphics[width=2.5cm]{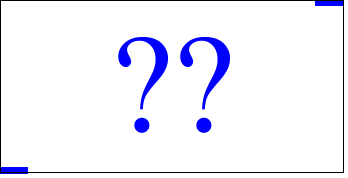}}	& & & \\ 
			& & & \\
			& & & \\
    \bottomrule
    \end{tabular}
    \end{small}
\end{table*}

\subsubsection{The interleaving attack}
Against the interleaving attack, the leading term of the code length does not depend on the choice of $p$ when the interleaving defense is used~\cite[Proposition 10]{oosterwijk13}, and its asymptotic value is always given by $\ell \sim 2 c^2 \ln n$. However, minimizing the variance of the coalition score by choosing $p = \frac{1}{2}$ leads to slightly smaller first order terms. In that case, the score function reduces to $+1$ for matches and $-1$ for differences.

\subsubsection{The all-$1$ attack}
For the all-$1$ attack (output a $1$ if at least one of the colluders received a $1$), the optimal choice of $p$ is given by $p = \frac{\alpha}{c}$, where $\alpha = o(1)$ should not converge to $0$ too quickly, to prevent first order terms from blowing up~\cite{laarhoven13allerton}. The all-$1$ attack is directly related to group testing~\cite{laarhoven13allerton,meerwald11b} where testing a pool of samples returns a positive result iff at least one positive item is present in the tested pool. Note that we have manually adjusted the function of \cite{oosterwijk13} from $h(1,0) = -1$ to $h(1,0) = -\infty$, since if a user has a $1$ and the pirate output is a $0$, he is certainly innocent. The asymptotic code length is given by $\ell \sim 2 c \ln n$ for this choice of $p$, while the simpler choice $p = \frac{1}{c}$ leads to $\ell \sim \frac{2e(e - 1)}{2e-1}c \ln n$.

Note that the all-$0$ attack is equivalent to the all-$1$ attack with the interpretation of the symbols $0$ and $1$ reversed. So we can obtain the optimal $h$ and $p$ by replacing each occurrence of $0$ by $1$ and $p$ by $1 - p$, and vice versa. 

\subsubsection{The coin-flip attack}
Similar to the all-$1$ attack, the optimal choice of $p$ against the coin-flip attack (randomly select either a $0$ or a $1$) is of the order $p = o(\frac{1}{c})$, leading to an asymptotic number of segments of $\ell \sim 4 c \ln n$ for large $n$. Again, smaller values of $p$ lead to larger first order terms, and so a compromise could be to take $p = \frac{1}{c}$ with an asymptotic code length of $\ell \sim 2(e^2 - 1) c \ln n$. 

\subsubsection{The majority voting attack}
For arbitrary values of $p$ and $c$, the optimal score function tailored against the majority voting attack (outputting the most common version among the pirate copies) looks quite nasty. However, optimizing $p$ leads to $p = \frac{1}{2}$, leading to the trivial but intuitive score function of $+1$ for matches and $-1$ for differences. Determining $\tilde{\mu}$ turns out to be related to the expected absolute distance from the origin after $c$ steps in a simple $1$-dimensional random walk, and working out the details leads to an asymptotic code length of $\ell \sim \pi c \ln n$.

\subsubsection{The minority voting attack}
Again, for arbitrary values of $p$ and $c$, the optimal score function built against the minority voting attack (always outputting the least common version among the pirate copies) does not look very pretty. However, the optimal choice of $p$ is close to $0$, in which case the minority voting attack is almost equivalent to the all-$1$ attack. The score function in the Table (which is the all-$1$ score function without manually optimizing $h$) is not exactly correct, but for reasonably large values of $c$ and small values of $p$, the difference is negligible. For $p = o(\frac{1}{c})$ the optimal scaling of $\ell$ is equivalent to that of the all-$1$ attack, while for $p = \frac{1}{c}$ it is slightly higher due to the manual optimization of the score function of the all-$1$ attack.

\subsubsection*{Other attacks}
Finally, for arbitrary other attacks not mentioned above, one can manually compute the optimal parameters as follows:
\begin{itemize}
	\item Compute the optimal score function $h$~\cite[Cor.~7~and~8]{oosterwijk13}.
	\item Determine the performance indicator ($\sigma^2/\tilde{\mu}^2$) as a function of $p$, and choose the value of $p$ that minimizes it.
\end{itemize}
Choosing $\ell$, $Z$, and $\delta$ as a function of $h$, $p$, and $c$ is not so easy, and as mentioned before depends on whether one wants provable security or only heuristic bounds on the error probabilities. Doing this analysis is left as an open problem.

\paragraph*{Remark 1} Looking at the examples above, we notice that the optimal choice of $p$ seems to be closely related to where the pirate strategy has the biggest jump in the probability of outputting a $1$ (cf. the graphs in Table~\ref{tab:1}). For instance, in the all-$1$ attack, this jump occurs between $\frac{k}{c} = 0$ and $\frac{k}{c} = \frac{1}{c}$, and the optimal choice of $p$ is indeed between $0$ and $\frac{1}{c}$. When the pirates use the majority voting attack, a jump only occurs when the fraction of ones goes from slightly below $\frac{1}{2}$ to slightly above $\frac{1}{2}$, and indeed the optimal choice of $p$ is exactly $\frac{1}{2}$. Note that except for against the majority attack, the asymptotic code length scales as $\ell \sim 2 c \ln (n) / \beta$, where $\beta$ is the size of the biggest jump. For instance, for the all-$1$ attack and minority attack, we have $\beta = 1$; for the coin-flip attack we have $\beta = \frac{1}{2}$; and for the interleaving attack we have $\beta = \frac{1}{c}$. This also illustrates why the interleaving attack is very strong.

\paragraph*{Remark 2} The above results can also be applied to the static traitor tracing setting to reduce the number of segments needed to catch colluders, but also to guarantee that \textit{all} colluders are caught with high probability. In fact, the above asymptotics for $\ell$ also hold for static traitor tracing if $c = n^{o(1)}$, although the convergence to the limit is considerably slower. For instance, the above parameters for the all-$1$ attack lead to improved non-adaptive (static) group testing schemes. For details, see~\cite{laarhoven13allerton}.


\subsection{Unknown attacks}

If the attack strategy is \textit{not} known to the distributor, he has to choose $h$ and the distribution function $F$ such that \textit{any} attack can be dealt with efficiently. An obvious suggestion would be to use the dynamic Tardos scheme with the score function replaced by the interleaving defense, as the interleaving defense is asymptotically optimal. However, we run into some practical issues when we try to do this, since the interleaving defense does not satisfy the same nice properties as the Tardos defense. If we use the Tardos score function, we know that $\sigma^2 = 1$, and so we can fix the threshold $Z$ in advance (based only on $c$, $n$, $\eps_1$, $\eps_2$) and get the proof methods of \cite{laarhoven13tit} to work. But if we use the interleaving defense, we only know that $\sigma^2/\tilde{\mu}^2 \leq 1$ for arbitrary attacks and large $c$, but we do not have any accurate estimates on $\sigma^2$. In fact, this parameter not only depends on the scheme parameters, but also on the attack strategy, which is unknown. For instance, using the minority attack leads to large values of both $\sigma$ and $\tilde{\mu}$, while the majority attack leads to small values for both. Thus, using a fixed threshold $Z$ does not seem to work.

\subsubsection{A heuristic construction}
We present a heuristic solution as follows, where the trick is that we vary $Z$ throughout, depending on $\sigma$. During the scheme, we keep track of the average innocent standard deviation per segment $\bar{\sigma}(i) = \sqrt{\frac{1}{i} \sum_{k \leq i} \sigma_k^2}$, where $\sigma_k^2 = p_k h(1,y_k,p_k)^2 + (1 - p_k) h(0,y_k,p_k)^2$ can be computed based on $p_k$ and $y_k$. Then, after each position $i$, instead of using $Z$ as a threshold, we use $Z(i) = Z \cdot \bar{\sigma}(i)$ as the threshold for disconnecting active users. The rest of the scheme remains the same.

\begin{figure*}[t]
	\centering
	\subfloat[][The Tardos defense vs. Arbitrary attacks]{\includegraphics[width=\columnwidth]{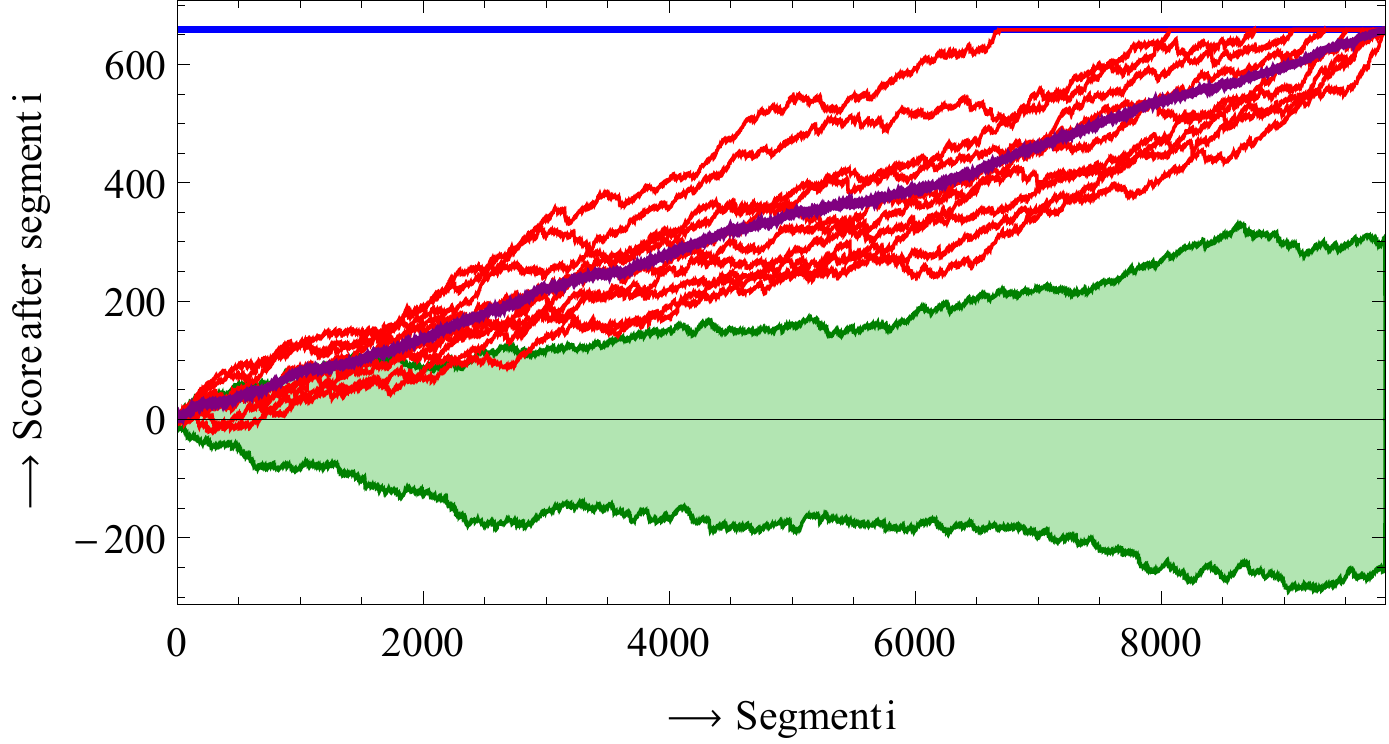} \label{fig:Fig1a}} \
	\subfloat[][The interleaving defense vs. The interleaving attack]{\includegraphics[width=\columnwidth]{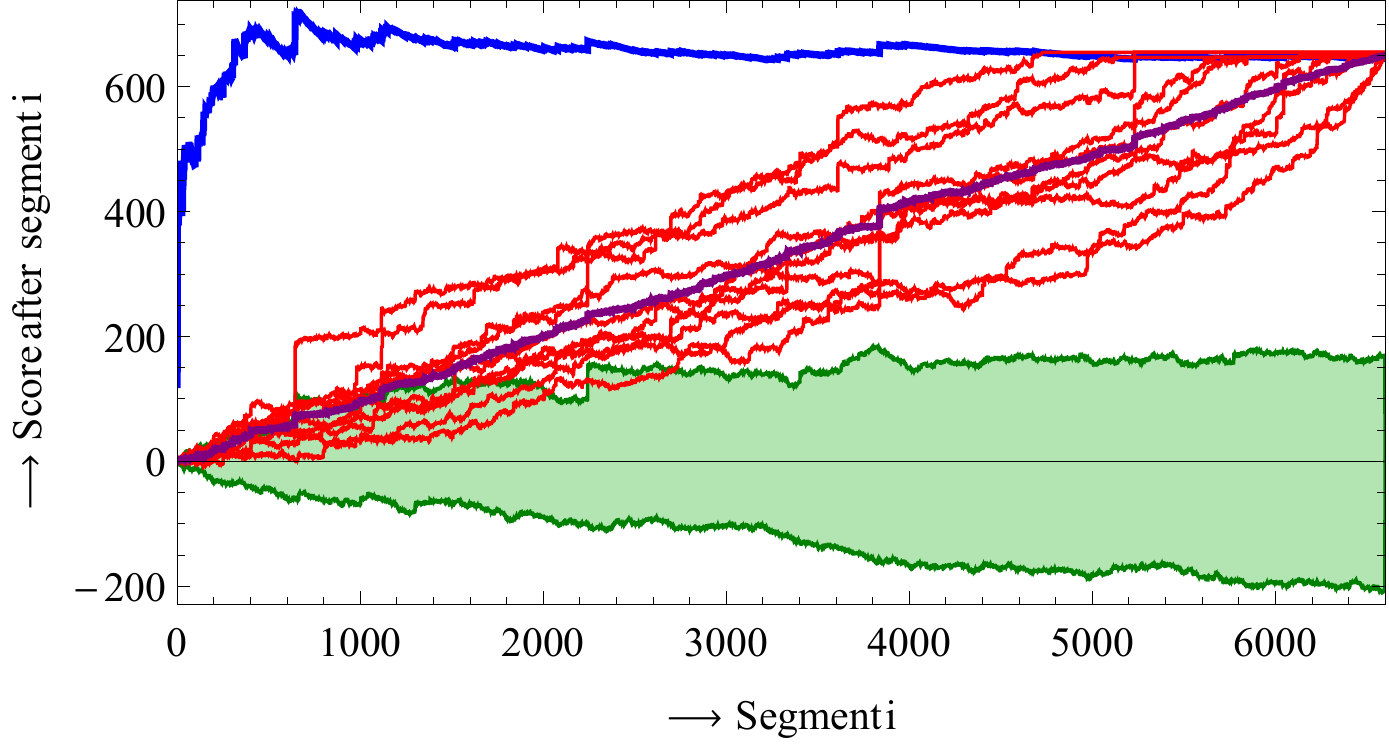} \label{fig:Fig1b}} \\
	\subfloat[][The interleaving defense vs. The majority attack]{\includegraphics[width=\columnwidth]{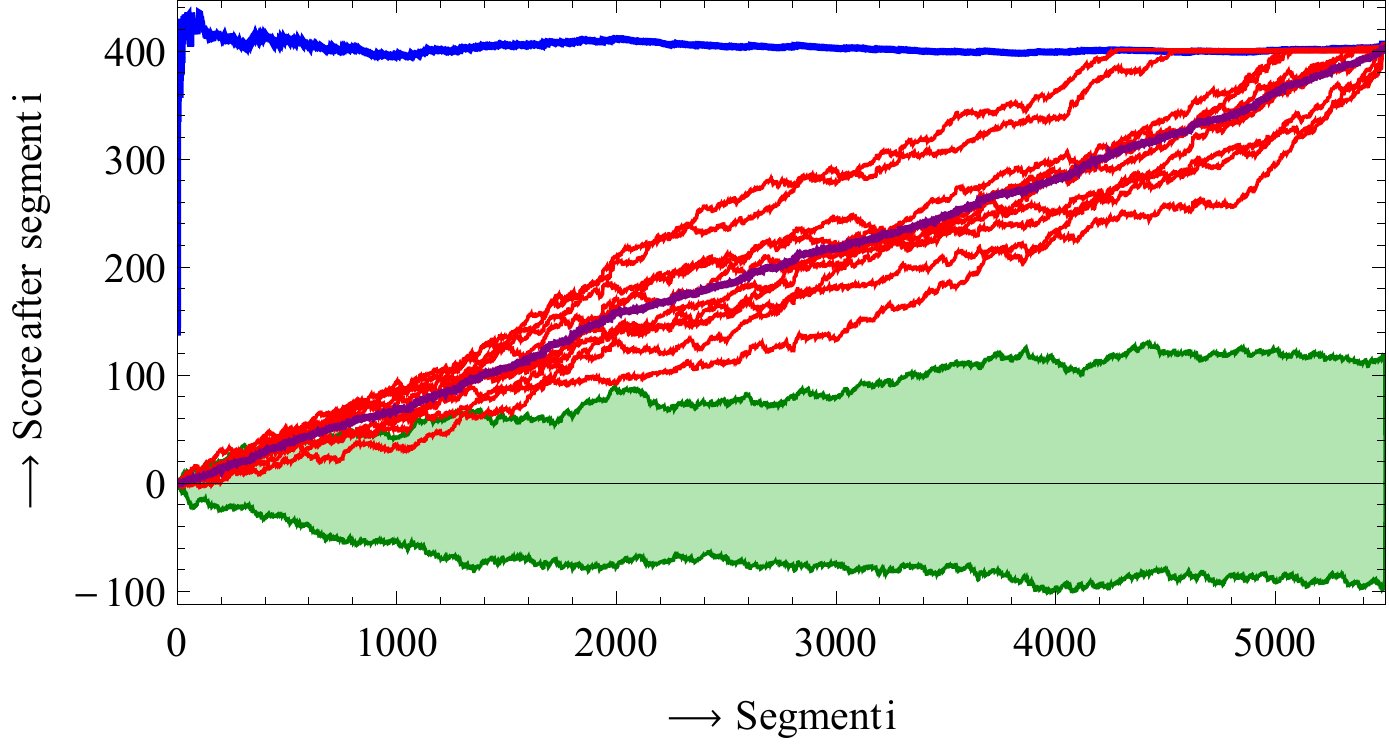} \label{fig:Fig1c}} \ 
	\subfloat[][The interleaving defense vs. The minority attack]{\includegraphics[width=\columnwidth]{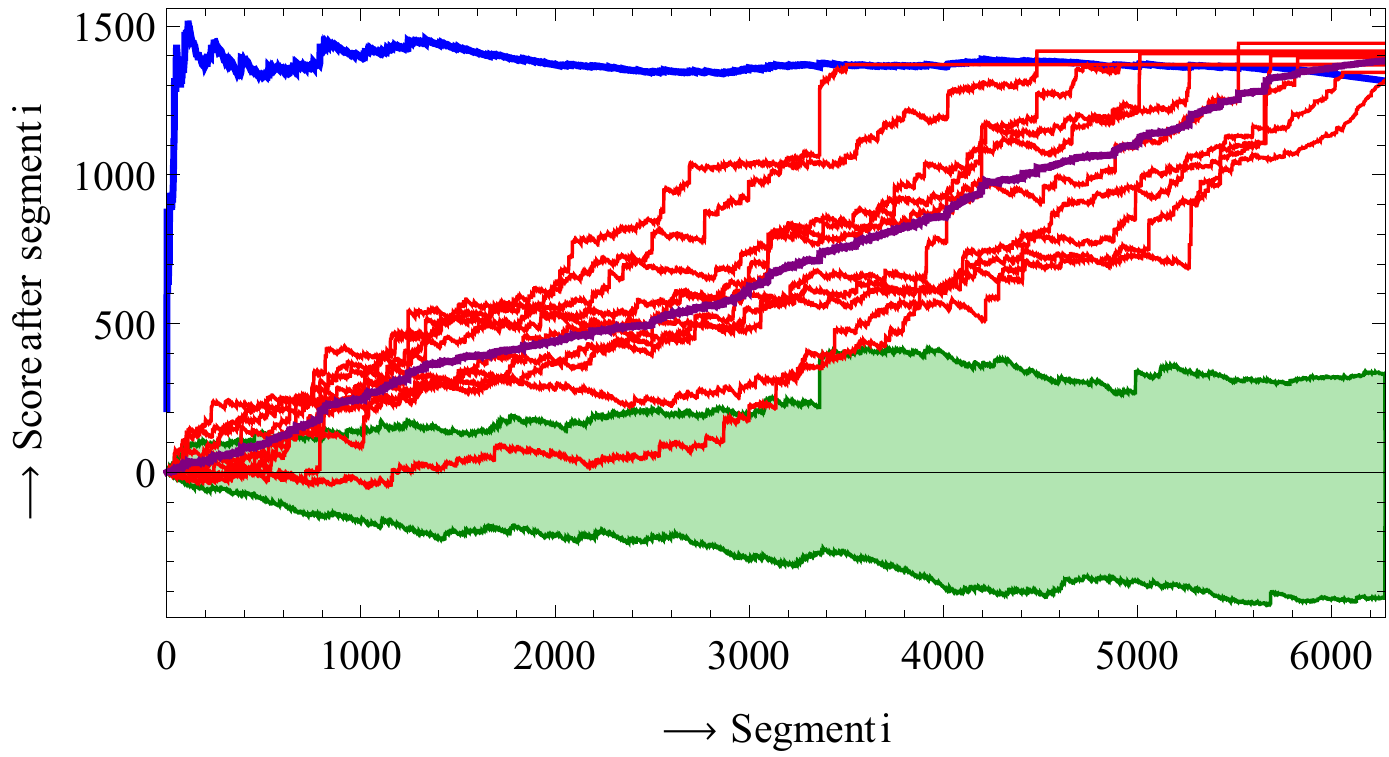} \label{fig:Fig1d}} \\
	\subfloat[][The interleaving defense vs. The maj/minority attack]{\includegraphics[width=\columnwidth]{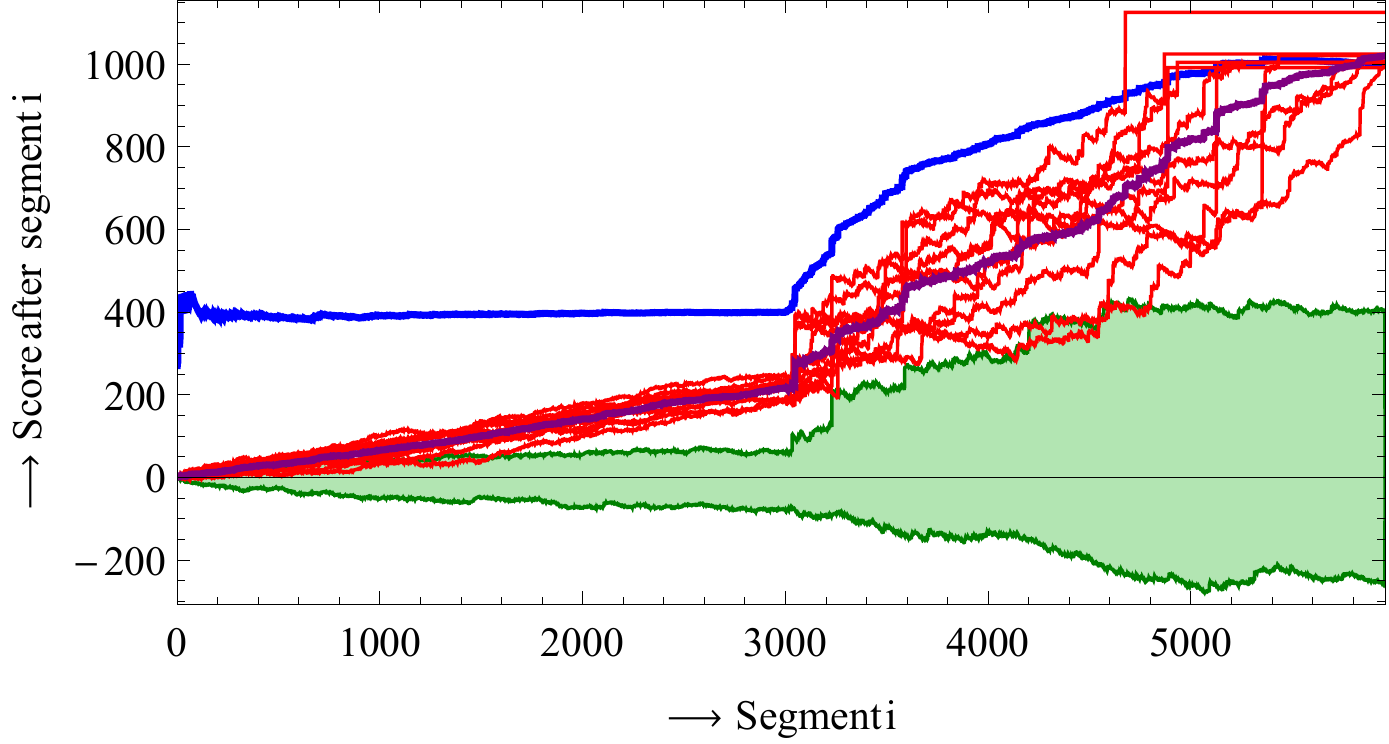} \label{fig:Fig1e}} \ 
	\subfloat[][The interleaving defense vs. The min/majority attack]{\includegraphics[width=\columnwidth]{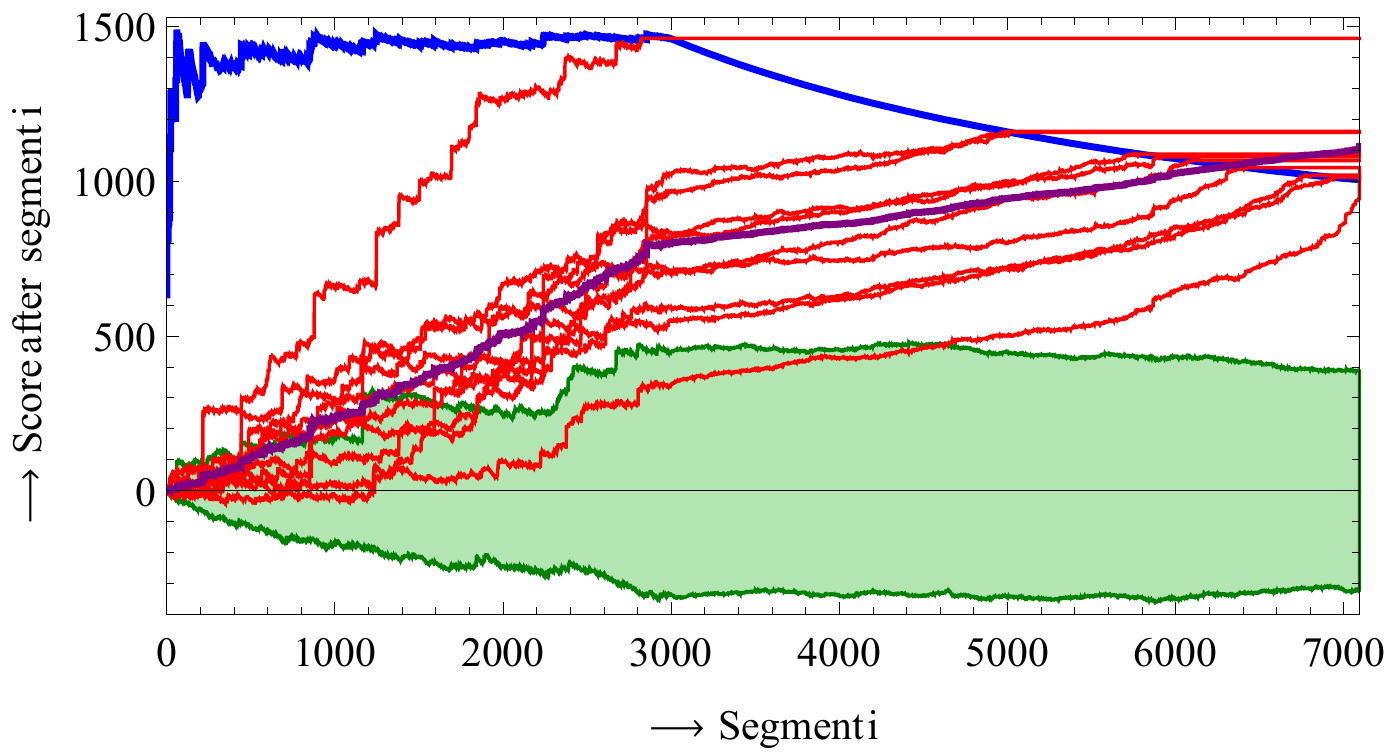} \label{fig:Fig1f}} \ 
	\caption{The dynamic interleaving defense in action against several attacks, together with one example of how the original dynamic Tardos scheme roughly works against arbitrary attacks. In this case, we used $c = 10$ and $n = 10^2$, $\eps_1 = \eps_2 = 10^{-3}$ as the simulation parameters, leading to $\ell \approx 14\,150$, $Z \approx 650$, and $\delta \approx 1/350$. The green area contains all innocent scores, the red lines show pirate scores, the bold purple line shows the average pirate score, and the bold blue curve shows the threshold $Z(i)$. The maj/minority attack consists of using the majority attack on the first $3000$ positions and then switching to the minority attack, while in the min/majority attack they start with the minority attack and end with the majority attack. Note that the numbers on both axes are different in each figure, so both the number of segments needed to catch all colluders and the order of magnitude of user scores are different in each example.}
	\label{fig:dynil}
\end{figure*}

In Figure~\ref{fig:dynil} we show some experiments using this construction, where we used the same parameters $\ell$, $Z$, and $\delta$ as in the dynamic Tardos scheme (which could be considered a conservative choice). Several attacks are considered with extreme values and fluctuations of $\sigma$, but in each case the scheme seems to work well, as all pirates are caught sooner than in the original dynamic Tardos scheme.  

\subsubsection{Possible issues}
Although the scheme seems to work well, we mention some theoretical issues with this scheme. 

\paragraph{The Gaussian assumption} 
Whereas in the Tardos scheme scores in each segment are identically distributed, the same does not hold for the interleaving defense. Some segments inevitably contribute much more to a user's overall score than other segments, due to varying values of $p$. Therefore, the assumption that the total user scores behave somewhat Gaussian may not hold, and the tails of the score curves may be bigger than expected. For instance, in Figures~\ref{fig:Fig1d}, \ref{fig:Fig1e}, and \ref{fig:Fig1f}, some positions cause both the green and the red curves to go up more that in other segments. Note that this issue exists for the interleaving defense in general, i.e., also for the interleaving defense applied to static traitor tracing.

\paragraph{Asymmetric attacks}
Related to the above, colluders may use column-asymmetric attacks (i.e., change their strategy over time) that cause large fluctuations in the parameters $\tilde{\mu}$ and $\sigma$ over time. An attack may for instance consist of first using the minority attack to cause $\sigma$ to go up, and then using the majority attack to make the value of $Z$ go down. This could possibly lead to innocent users being disconnected, because of their initially high scores. Figures~\ref{fig:Fig1e}, \ref{fig:Fig1f} suggest that such attacks are not very effective, but again, simulations do not guarantee that there is no other way to exploit this issue.

In both cases, choosing the cutoff $\delta$ sufficiently large seems crucial. In the original Tardos scheme things may already go wrong if the cutoff is too small, but in the case of the interleaving defense, the effects of small values of $p$ are much more extreme. Not using a cutoff may lead to explosions of user scores when $p$ happens to be very small and the pirates still manage to output a $1$, and in general a too small cutoff may cause huge fluctuations in $\sigma$ and $\tilde{\mu}$ in different segments. But if we do use an appropriate cutoff, as we did in Figure~\ref{fig:dynil}, these issues might not cause any practical problems at all.

Finally, note that if we were to normalize the scores in each segment, to avoid the two issues mentioned above, we end up with the Tardos score function again. So using the Tardos score function may still be a practical choice to guarantee theoretical security and avoid these possible issues mentioned above. But heuristically, using the interleaving defense seems pretty safe, and can lead to drastic reductions in the code length.


\section{Conclusion}
\label{sec:conclusion}

\noindent To summarize the most important contributions of this paper:
\begin{itemize}
	\item By making a small modification to the construction of the dynamic Tardos scheme, we get a simplified proof and a slightly improved performance.
	\item Against known attacks, optimizing $p$ and $h$ may lead to much faster tracing of collusions. The asymptotic code length often decreases from quadratic to linear in $c$.
	\item Against unknown attacks, we can use the interleaving defense and the arcsine distribution with cutoffs to catch arbitrary collusions much faster, by varying $Z$ throughout. 
	\item Some theoretical issues exist with the interleaving defense, but experimentally these issues seem less relevant. However, using sufficiently large cutoffs seems crucial.
\end{itemize}
An open problem remains whether the interleaving defense can provably withstand arbitrary attacks. For now, it seems that using the interleaving defense is quite safe and very efficient.



\section*{Acknowledgment}

The author is grateful to Jan-Jaap Oosterwijk and Benne de Weger for their valuable comments and suggestions.


\bibliographystyle{IEEEtran}

\bibliography{IEEEabrv,wifs13-refs}

\end{document}